\documentclass[draft]{agujournal2019}

\usepackage{url} 
\usepackage{amsmath,amssymb}
\usepackage[inline]{trackchanges} 
\usepackage{soul}
\usepackage{setspace}
\usepackage{microtype}
\usepackage{silence}
\microtypesetup{expansion=false}

\newcommand{\citeg}[1]{\cite<e.g.,>[]{#1}}
\newcommand{\citef}[1]{\cite<c.f.,>[]{#1}}

\draftfalse

\newcommand{\Grad}{\boldsymbol{\nabla}}

\newcommand{\pdiff}[2]{\frac{\partial{#1}}{\partial{#2}}}

\newcommand{\RNum}[1]{\uppercase\expandafter{\romannumeral #1\relax}}

\newcommand{\secref}[1]{Section \ref{#1}}
\newcommand{\figref}[1]{Figure \ref{#1}}

\newcommand{\tabref}[1]{Table \ref{#1}}
\newcommand{\appref}[1]{\ref{#1}}
\newcommand{\eqnref}[1]{Equation \ref{#1}}

\journalname{JGR: Earth Surface}

\begin{document}

\title{A Unified Blister and Subglacial Hydrology Framework for Supraglacial Lake Drainage Events}

\authors{Hanwen Zhang\affil{1}, Laura A.\ Stevens\affil{1}, Ian J.\ Hewitt\affil{2}, Harry Stuart\affil{2}}

\affiliation{1}{Department of Earth Sciences, University of Oxford, South Parks Road, Oxford OX1 3AN, UK}
\affiliation{2}{Mathematical Institute, University of Oxford, Woodstock Road, Oxford OX2 6GG, UK}

\correspondingauthor{Hanwen Zhang}{hwenzhang.ac@gmail.com}

\begin{keypoints}

\item Blister dynamics incorporated into a 2D subglacial hydrology model to explore the transient effects of rapid supraglacial lake drainages.

\item Two competing factors control blister dynamics: lateral propagation and leakage into the surrounding drainage system.

\item Simulations of a lake drainage event in Greenland show trajectories and speeds consistent with remote-sensing observations.

\end{keypoints}

\begin{abstract}
Subglacial blisters form due to the rapid drainage of supraglacial lakes into grounded ice sheets, and are characterised by elastic ice uplift and transient ice-velocity anomalies. Although blister occurrence is confirmed by observations, the dynamics of blisters and their impacts on ice flow remain poorly represented in current subglacial hydrology models, as typical cavity-channel system models cannot capture short-timescale blister formation, propagation, and relaxation. Here we present a unified, self-consistent modelling framework that directly couples blister evolution with the subglacial drainage system, extending existing subglacial hydrology models to account for transient responses to rapid lake drainage events. Numerical simulations, motivated by field observations of wintertime lake drainages, reveal distinct seasonal behavior: during summer, lake drainage generates short-lived blisters that rapidly leak water into a pre-existing drainage system of efficient, channelised water pathways, whereas winter drainage results in persistent blisters that propagate slowly and serve as the primary meltwater pathway at the ice-bed interface. The dynamics of blister propagation and leakage in our model are governed by effective viscosity and a characteristic leakage length scale, which reflects the connection between the blister and the surrounding hydrological network. This unified model offers a valuable tool for investigating blister dynamics and their interplay with subglacial hydrology, facilitating the interpretation of observed surface uplift and ice-velocity variations following supraglacial lake drainage events.
\end{abstract}

\section*{Plain Language Summary}
Supraglacial lakes, which form on the surface of ice sheets, can rapidly drain their water to the bed through vertical cracks in the ice sheet. This sudden water influx to the ice--bedrock interface can cause the ice to lift up, creating a so-called ``blister", which will in turn affect water flow on the ice--bedrock interface and the overlying ice movement. However, current models do not account for these blisters. In this study, we develop a new model that couples blister dynamics with the water flow beneath the ice. Our results show how blisters form, propagate, and leak water into the surrounding drainage system, with differences in model behaviour depending on the season. This new model helps us better understand how blisters affect ice movement and water flow beneath ice sheets, which is important for predicting future changes in how ice sheets flow and contribute to sea level rise.

\section{Introduction}\label{sec:intro}
\begin{figure}[!ht]
    \centering
    \includegraphics[width=0.8\linewidth]{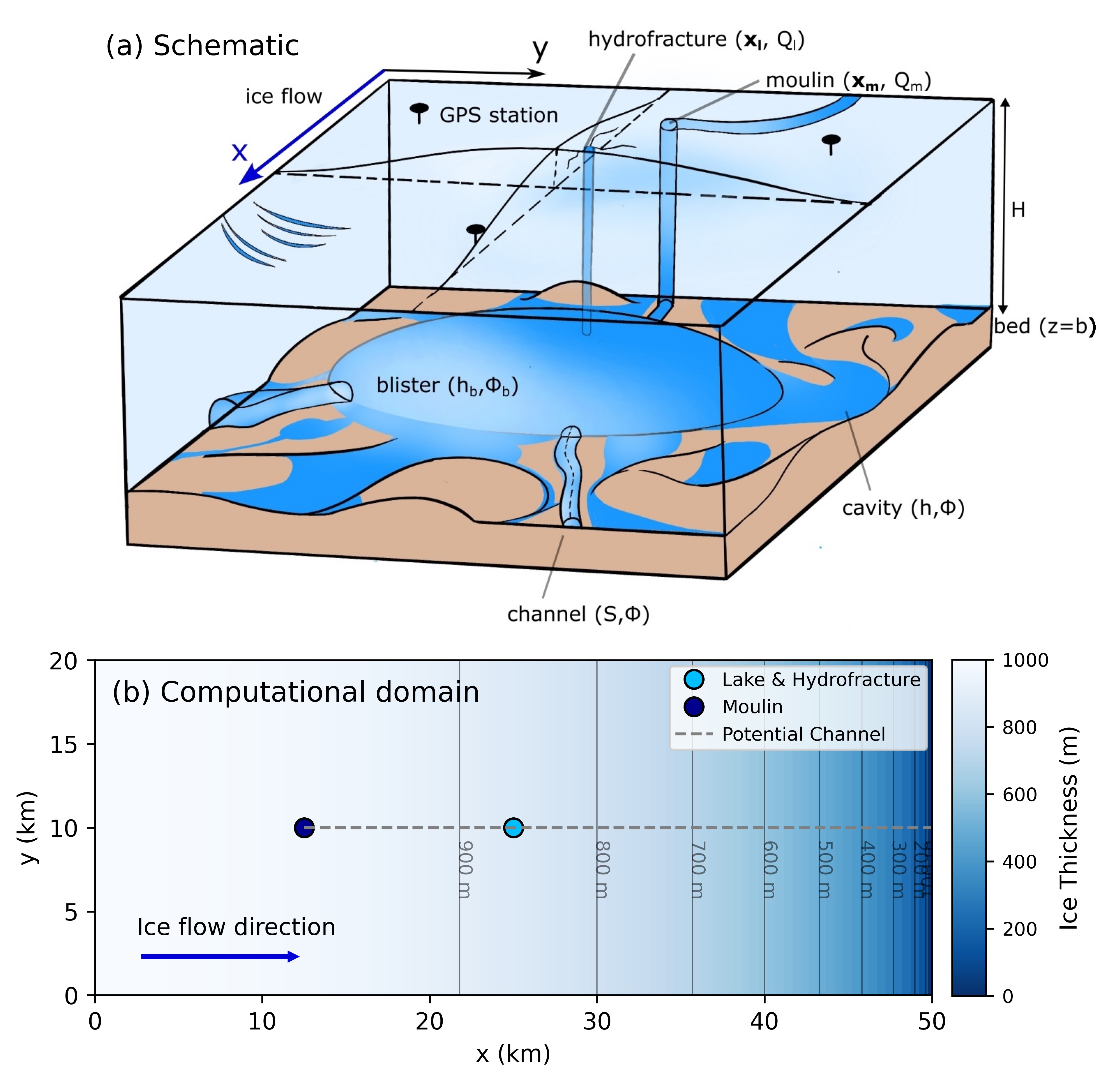}
    \caption{\textbf{Unified model set-up and computational domain.} (\textbf{a}) A schematic of the subglacial hydrology model including blister, cavity and channel components. (\textbf{b}) The computational domain for the reference cases shown in Figures~\ref{fig:wintertime_drainage} and~\ref{fig:summertime_drainage}.}
    \label{fig:schematic}
\end{figure}

On the Greenland Ice Sheet, rapid drainage of supraglacial lakes via hydrofracture can transport substantial volumes of meltwater into the subglacial drainage system within hours. Such drainage events have been widely observed near Greenland Ice Sheet margins in both summer \cite{das2008Fracture,stevens2026Hydrofracture} and winter \cite{maier2023Wintertime,dean2026Decadea}, and can affect the seasonal evolution of subglacial hydrology and ice dynamics \cite{andrews2018Seasonal}. When lake-drainage input exceeds the capacity of the local subglacial drainage system, this input can generate elastic uplift of the overlying ice \cite{das2008Fracture,doyle2013Ice,stevens2015Greenland}, forming subglacial “blisters”(\figref{fig:schematic}a) that propagate laterally \citeg{lister2013Viscous} and leak water into the local subglacial drainage network \citeg{lai2021Hydraulic}. These processes can also cause transient speed-up of the overlying ice sheet \cite{doyle2013Ice,maier2023Wintertime}. Blisters, therefore, are a key hydrological component of the subglacial system, linking rapid surface meltwater delivery to the bed, short-lived basal water storage, and the transient response of the subglacial drainage system and overlying ice sheet during the rapid drainage of supraglacial lakes.

The effects of lake-drainage events and their associated blisters depend on the existing state of the local subglacial drainage system \cite{lai2021Hydraulic}. In winter, the drainage system is typically inefficient and dominated by water flow through linked cavities \cite{schoof2010Icesheet}, such that lake-drainage-induced surface velocity waves and uplift signals can propagate over tens of kilometres \cite{maier2023Wintertime}. In summer, the drainage system is more efficient and dominated by channels \cite{schoof2010Icesheet}, such that blisters tend to leak rapidly into the channel network and decay quickly \cite{stevens2022Tidewaterglacier}. Understanding how blisters evolve and interact with the subglacial drainage system is crucial for characterising the transient ice-flow variations triggered by supraglacial lake-drainage events, and for extending our knowledge of subglacial hydrology evolution from seasonal to sub-daily timescales.

The dynamics of subglacial blisters have received considerable modelling interest. \citeA{tsai2010Model,tsai2012Modeling} treated blister formation as planar hydrofracture occurring near a free surface, and applied linear elastic fracture mechanics coupled with lubrication theory to estimate the velocity of the advancing blister front. An alternative approach models the blister as viscous water flow beneath an elastically bending ice sheet \cite{lister2013Viscous,hewitt2015Elasticplated,hewitt2018Influence,lister2019Viscous,peng2020Viscous,tobin2023Evolution}. Under certain idealised conditions, these studies show that the thickness of spreading blisters follows a power-law relationship with time. Finally, \citeA{lai2021Hydraulic} investigated blister relaxation due to leak-off into a porous substrate in a toughness-dominated regime, where radial blister expansion is negligible compared with leakage into the local drainage system. Their findings indicate that the thickness of such blisters decays exponentially over time.

Despite the insights gained from these pure blister models, effectively integrating blister dynamics into existing subglacial hydrology models remains challenging. Subglacial hydrology models typically consist of drainage components such as channels, cavities, and porous layers \citeg{hewitt2013Seasonal,werder2013Modeling,sommers2018SHAKTI,kazmierczak2024Fast}, but they do not explicitly capture blister formation, propagation, or relaxation on the short timescales associated with rapid lake-drainage events \cite{flowers2015Modelling}. Consequently, these models have limited ability to accommodate intense, transient water inputs or to accurately represent blister dynamics. To address these limitations, \citeA{pimentel2010Numerical} modelled blisters as elastic ice flexures interacting with a subglacial drainage system represented by an elastic porous sheet. Alternatively, \citeA{rice2015Time} and \citeA{dow2015Modeling} combined the hydrofracture model developed by \citeA{tsai2012Modeling} with a one-dimensional subglacial hydrology model, using hydrofracture solutions as initial conditions for subsequent hydrological evolution. However, both approaches remain one-dimensional and depend on simplifying assumptions regarding the prevailing subglacial drainage system, limiting their ability to fully resolve blister formation and propagation.

In this study, we propose a new modelling framework that accounts for the formation and propagation of two-dimensional blisters and their interaction with the evolving subglacial drainage system. Building on the approach that models blisters as viscous water flow beneath an elastically bending ice sheet \citeg{tobin2023Evolution}, we incorporate the blister model into a two-dimensional subglacial hydrology model \cite{hewitt2013Seasonal} by introducing a leakage term that links the blister to the local subglacial drainage system. Our model expands the capabilities of subglacial hydrology models by capturing the transient impacts of lake drainage events, enabling an integrated exploration of blister evolution and subglacial floods. We apply our framework to a land-terminating ice-sheet region in western Greenland, providing insights into previously observed surface uplift and velocity changes triggered by wintertime supraglacial lake drainage events \cite{maier2023Wintertime}.

The paper is organised as follows. \secref{sec:method} introduces the mathematical formulation of the blister model and its coupling with the subglacial drainage system. \secref{sec:results} describes the model set-up and illustrates two reference scenarios for wintertime and summertime lake drainage events. Parameter controls on blister propagation and leakage are explored in \secref{sec:discussion}, followed by a regional case study of a land-terminating ice-sheet region in western Greenland (\secref{sec:regional_study}), demonstrating the model's ability to be applied to study subglacial floods resulting from multiple wintertime lake drainage events. Finally, we summarise our findings and discuss model limitations and future directions in Sections~\ref{sec:limitations} and~\ref{sec:conclusion}.

\section{Method}\label{sec:method}
The unified model consists of two coupled water systems: a background cavity-channel drainage system and a transient blister layer (\figref{fig:schematic}a). The cavity-channel system describes water routed through linked cavities and R-channels, whereas the blister layer, located on top of the cavity-channel system, describes water flowing beneath an elastically flexing ice sheet after rapid lake drainage. The two systems interact through a leakage term, so that water lost from the blister is gained by the cavity-channel system.

\subsection{Cavity-channel system}\label{sec:method_cavity_channel}
The cavity-channel component follows the two-component subglacial hydrology model of \citeA{hewitt2013Seasonal} and \citeA{werder2013Modeling}. The unknowns are the cavity sheet thickness $h$, the hydraulic potential $\phi$, and the channel cross-sectional area $S$. The cavity sheet and channels share
\begin{linenomath*}
\begin{equation}\label{eq:hydraulic_potential_cavity}
    \phi = \rho_w g b + p_w,
    \qquad
    N = p_i-p_w = \rho_i g H-p_w,
\end{equation}
\end{linenomath*}
where $p_w$ is the water pressure in the cavity-channel system, $N$ is effective pressure, $p_i=\rho_i gH$ is ice overburden, $b$ is bed elevation, and $H$ is ice thickness. The sheet flux is written in the general power-law form used in the numerical model,
\begin{linenomath*}
\begin{equation}\label{eq:sheet_flux}
    \boldsymbol{q}
    =-K_s h^{\alpha_s}\left|\Grad\phi\right|^{\beta_s-1}\Grad\phi,
\end{equation}
\end{linenomath*}
with $\alpha_s=3$ and $\beta_s=1$ in the reference simulations. Channel flux along the channel coordinate $s$ is
\begin{linenomath*}
\begin{equation}\label{eq:channel_flux}
    Q
    =-K_c S^{\alpha_c}\left|\pdiff{\phi}{s}\right|^{\beta_c-1}\pdiff{\phi}{s},
\end{equation}
\end{linenomath*}
with $\alpha_c=5/4$ and $\beta_c=1/2$. These flux laws route water down gradients of hydraulic potential in the sheet and channel networks, respectively.

The cavity opening and closure relation is
\begin{linenomath*}
\begin{equation}\label{eq:cavity_evolution}
    \pdiff{h}{t}
    =
    \frac{\rho_w}{\rho_i}m
    + U_b\frac{h_r-h}{l_r}
    - \frac{2A}{n^n}h|N|^{n-1}N,
\end{equation}
\end{linenomath*}
where the three terms represent melt opening, sliding opening over bed roughness, and viscous creep closure. The basal sliding speed $U_b$ is prescribed in the simulations presented here. The channel area evolves as
\begin{linenomath*}
\begin{equation}\label{eq:channel_evolution}
    \pdiff{S}{t}
    =
    \frac{\rho_w}{\rho_i}M
    - \frac{2A}{n^n}S|N|^{n-1}N,
\end{equation}
\end{linenomath*}
where channel opening is caused by dissipative melting and channel closure is caused by ice creep. We parameterise the channel melting rate as
\begin{linenomath*}
\begin{equation}\label{eq:channel_melting_method}
    M = \frac{\left|Q\,\partial\phi/\partial s\right|
    + \lambda_c\left|\boldsymbol{q}\cdot\Grad\phi\right|}
    {\rho_w \mathcal{L}},
\end{equation}
\end{linenomath*}
where $\lambda_c$ is the width over which dissipation in the surrounding sheet contributes to channel melt, and $\mathcal{L}$ is the latent heat of melting.

Mass conservation for the combined cavity sheet and channel storage is expressed as
\begin{linenomath*}
\begin{equation}\label{eq:mass_conservation}
    \pdiff{h}{t}
    + \nabla \cdot \boldsymbol{q}
    + \left(\pdiff{S}{t}+\pdiff{Q}{s}\right)\delta(\boldsymbol{x}_c)
    = m + M\delta(\boldsymbol{x}_c) + \Sigma_m + Q_b,
\end{equation}
\end{linenomath*}
where $\boldsymbol{x}_c$ denotes the channel position, $\delta(\boldsymbol{x}_c)$ is the Delta function that localises the channel terms, $m$ is basal melt entering the sheet, and $M$ is channel-wall melt. The moulin input term, $\Sigma_m$, represents the point-source surface input to the cavity-channel system through moulins,
\begin{linenomath*}
\begin{equation}\label{eq:moulin_cavity}
    \Sigma_m
    =
    \sum_m Q_m \delta(\boldsymbol{x}_m)\mathcal{H}\left(N(\boldsymbol{x}_m)\right),
\end{equation}
\end{linenomath*}
where $\boldsymbol{x}_m$ is the moulin position and $\mathcal{H}$ is a smoothed Heaviside function in the numerical implementation. This switch sends moulin water to the cavity-channel system when the local effective pressure is non-negative, and a corresponding term in (\eqnref{eq:mass_conservation_blister}) below will divert the input to the blister when the cavity-channel system is overwhelmed and $N<0$.  
The last term, $Q_b$, is the leakage term that represents the leakage rate from the blister layer to the cavity-channel system, and will be defined in \secref{sec:method_blister}.


\subsection{Blister model}\label{sec:method_blister}
The blister component describes a water layer of thickness $h_b$ beneath an elastically flexing ice sheet. Its hydraulic potential is defined as
\begin{linenomath*}
\begin{equation}\label{eq:hydraulic_potential_blister}
    \phi_b = \rho_w g b + p_b,
    \qquad
    p_b = \rho_i g H + \rho_w g h_b
    + \nabla^2 \left(B\nabla^2 h_b\right),
\end{equation}
\end{linenomath*}
where $p_b$ is blister water pressure. The three contributions to $p_b$ are ice overburden, the hydrostatic pressure associated with the blister thickness, and elastic bending of the overlying ice. The bending stiffness is $B=E H^3/[12(1-\nu^2)]$, where $E$ is Young's modulus and $\nu$ is Poisson's ratio. \eqnref{eq:hydraulic_potential_blister} is derived from the vertical force balance on an elastically bending ice sheet overlying the blister. The blister water pressure must support the ice overburden, the hydrostatic pressure associated with the water layer, and the elastic load required to bend the ice.

Following lubrication theory, we write the water flux within the blister as
\begin{linenomath*}
\begin{equation}\label{eq:blister_flux}
    \boldsymbol{q}_b
    =
    -\frac{(h_b+h_0)^3}{12\mu_{\text{eff}}}\Grad\phi_b,
\end{equation}
\end{linenomath*}
where $\mu_{\text{eff}}$ is an effective viscosity. This effective viscosity is used to account for turbulence, ice-bed roughness and other characteristics of the ice-bed interface that contribute to energy loss during propagation. In reality, this resistance may vary in space and time as the blister evolves, depending on local bed conditions, water pressure, flow speed, and hydraulic connectivity with the surrounding drainage system. Alternative formulations have also been used to represent such nonlinear and heterogeneous flow behaviour \cite<e.g.>[]{zimmerman2004non,sommers2018SHAKTI,hill2024Improved}, but we treat $\mu_{\text{eff}}$ as a constant to isolate its first-order control on blister propagation and keep the reference simulations readily interpretable.

The regularisation parameter $h_0$ mimics a pre-wetted layer ahead of the propagating front \cite{hewitt2015Elasticplated}, a common treatment in blister models \citeg{hewitt2018Influence,tobin2023Evolution}. Although $h_0$ influences the propagation velocity, the dependence is weak \cite{peng2020Viscous}. We use $h_0=0.005$ m in the reference cases.

The blister mass balance is
\begin{linenomath*}
\begin{equation}\label{eq:mass_conservation_blister}
    \pdiff{h_b}{t}
    + \nabla\cdot\boldsymbol{q}_b
    =
    Q_l + \Sigma_m^{b} - Q_b,
\end{equation}
\end{linenomath*}
where $Q_l$ is lake-drainage input, $\Sigma_m^{b}$ is the moulin input diverted into the blister, and $Q_b$ is the leakage from the blister into the cavity-channel system. For the reference cases, lake drainage is imposed at $\boldsymbol{x}_l$ as a Gaussian pulse,
\begin{linenomath*}
\begin{equation}\label{eq:lake_input}
    Q_l
    =
    \frac{V_l}{\sqrt{2\pi}\tau_l}
    \exp\left[-\frac{(t-t_l)^2}{2\tau_l^2}\right]
    \delta\left(\boldsymbol{x}_l\right),
\end{equation}
\end{linenomath*}
where $V_l$ is lake volume, $\tau_l$ is drainage timescale, and $t_l$ is the drainage time. We use $\tau_l=0.6$ hr so that drainage occurs within hours, consistent with observations of lake hydrofracture events \cite{das2008Fracture,doyle2013Ice,stevens2015Greenland,chudley2019lake}.

The diverted moulin input is, mirroring (\ref{eq:moulin_cavity}),
\begin{linenomath*}
\begin{equation}\label{eq:moulin_blister}
    \Sigma_m^{b}
    =
    \sum_m Q_m \delta(\boldsymbol{x}_m)
    \left[1-\mathcal{H}\left(N(\boldsymbol{x}_m)\right)\right].
\end{equation}
\end{linenomath*}
This means that the water from each moulin enters either the blister layer or the cavity-channel system depending on local effective pressure.

For the leakage term $Q_{\text{b}}$ that links the blister and the surrounding drainage system, we assume that leakage is proportional to the difference in hydraulic potential from the blister to the cavity-channel system:
\begin{linenomath*}
\begin{equation}\label{eq:leakage}
    Q_b = \frac{l\left(h_b,h,S\right)}{\mu_{\text{eff}}}\langle\phi_b-\phi\rangle,
\end{equation}
\end{linenomath*}
where $l$ has dimensions of length and is interpreted as a lengthscale that parameterizes unresolved small-scale controls on blister-to-drainage-system connectivity, including bed roughness, local cavity opening, connection density, and the transmissivity of nearby cavities and channels. Larger values of $l$ therefore correspond to stronger hydraulic connection and more rapid leakage from the blister, whereas smaller values correspond to a more isolated blister. For simplicity, we assume $l=\kappa h_b$, where $\kappa$ is a dimensionless connectivity coefficient, so that leakage increases as the blister opens. The Macaulay brackets $\langle \cdot \rangle$ denote the positive part, $\langle x\rangle=\max(x,0)$, ensuring that water leaks from the blister into the drainage system only when $\phi_b>\phi$. 

\section{Results}\label{sec:results}
To illustrate our model set-up, we follow \citeA{hewitt2013Seasonal} to construct an idealised land-terminating ice sheet (Fig.~\ref{fig:schematic}b), with a length $L$ and a width $W$. The ice sheet sits on a flat bed with an elevation $b=0$. The ice thickness, denoted as $H$, is a function of along-flow coordinate $x$:
\begin{linenomath*}
\begin{equation}
    H = H_0\sqrt{1-\frac{x}{L}},\quad 0<x<L,
\end{equation}
\end{linenomath*}
where $H_0$ is a constant.

We are interested in how the blister propagates and leaks into the local drainage system. Meanwhile, we explore the seasonality of blister dynamics, specifically how blister propagation differs between summer and winter. For simplicity, we assume that there is one moulin with influx $Q_m$ at $\boldsymbol{x}_m=(0.25L,0.5W)$ and one downstream supraglacial lake at $\boldsymbol{x}_l=(0.5L,0.5W)$. The subglacial drainage system is initialised using two different moulin influxes $Q_m=0~\text{m}^3~\text{s}^{-1}$~\text{and}~$Q_m=10~\text{m}^3~\text{s}^{-1}$, which represent the wintertime and summertime surface runoff, respectively. After the drainage system reaches a steady state driven by $Q_m$, a lake-drainage input $Q_l$ is imposed at $\boldsymbol{x}_l$, delivering $V_l=10^{7}~\text{m}^3$ of water into the blister within $\sim3$ hr, after which $Q_l$ returns to zero. The drainage event triggers the formation and propagation of a blister and consequent evolution of the subglacial drainage system. Reference values for the physical variables are given in \tabref{tab:parameters}.

\begin{table}[htbp]
    \centering
    \caption{\textbf{Parameter values.} Values in parentheses are used in the regional studies in \secref{sec:regional_study} if different from the reference cases.}
    \begin{tabular}{lcc}
        \hline
        Parameter & Symbol & Value \\
        \hline
        \multicolumn{3}{l}{\textit{Blister model parameters}} \\
        \hline
        Ice density & $\rho_i$ & $910~\text{kg}\cdot\text{m}^{-3}$ \\
        Water density & $\rho_w$ & $1000~\text{kg}\cdot\text{m}^{-3}$ \\
        Gravitational acceleration & $g$ & $9.81~\text{m}\cdot\text{s}^{-2}$ \\
        Ice thickness & $H_0$ & $1000~\text{m}$ \\
        Ice length & $L$ & $50~\text{km}$ \\
        Ice width & $W$ & $20~\text{km}$ \\
        Water viscosity & $\mu_{\text{eff}}$ & $10^{1}~\text{Pa}\cdot\text{s}$ \\
        Pre-wetted layer thickness & $h_0$ & $0.005~\text{m}$ \\
        Moulin influx & $Q_m$ & $0~\text{and}~10~\text{m}^3\cdot\text{s}^{-1}$ \\
        Lake drainage volume & $V_l$ & $10^{7}~\text{m}^3$ \\
        Lake drainage timescale & $\tau_l$ & $0.6~\text{hr}$ \\
        Young's modulus of ice & $E$ & $9.0\times10^{9}~\text{Pa}$ \\
        Poisson's ratio of ice & $\nu$ & $0.3$ \\
        Leakage coefficient & $\kappa$ & $10^{-10}$ \\
        \hline
        \multicolumn{3}{l}{\textit{Cavity-sheet and channel parameters}} \\
        \hline
        Latent heat of melting & $\mathcal{L}$ & $3.35\times10^{5}~\text{J}\cdot\text{kg}^{-1}$ \\
        Geothermal heat flux & $G$ & $0.063~\text{W}\cdot\text{m}^{-2}$ \\
        Ice flow law exponent & $n$ & $3$ \\
        Ice flow law coefficient & $A$ & $6.8\times10^{-24}~\text{Pa}^{-3}\cdot\text{s}^{-1}$ \\
        Sheet flux coefficient & $K_s$ & $10^{-4}~(10^{-3})~\text{Pa}^{-1}\cdot\text{s}^{-1}$ \\
        Sliding velocity & $U_b$ & $0~(100)~\text{m}\cdot\text{yr}^{-1}$ \\
        Basal friction & $\tau_b$ & $0~(60)~\text{kPa}$ \\
        Sheet roughness height & $h_r$ & $0.1~\text{m}$ \\
        Sheet roughness length & $l_r$ & $10~\text{m}$ \\
        Channel flux coefficient & $K_c$ & $0.1~\text{m}\cdot\text{s}^{-1}\cdot\text{Pa}^{-1/2}$ \\
        Sheet width contributing to channel melting & $\lambda_c$ & $10~(1000)~\text{m}$ \\
        \hline
    \end{tabular}
    \label{tab:parameters}
\end{table}

\subsection{Wintertime lake drainage}\label{sec:wintertime}
\begin{figure}[!ht]
    \centering
    \includegraphics[width=0.8\linewidth]{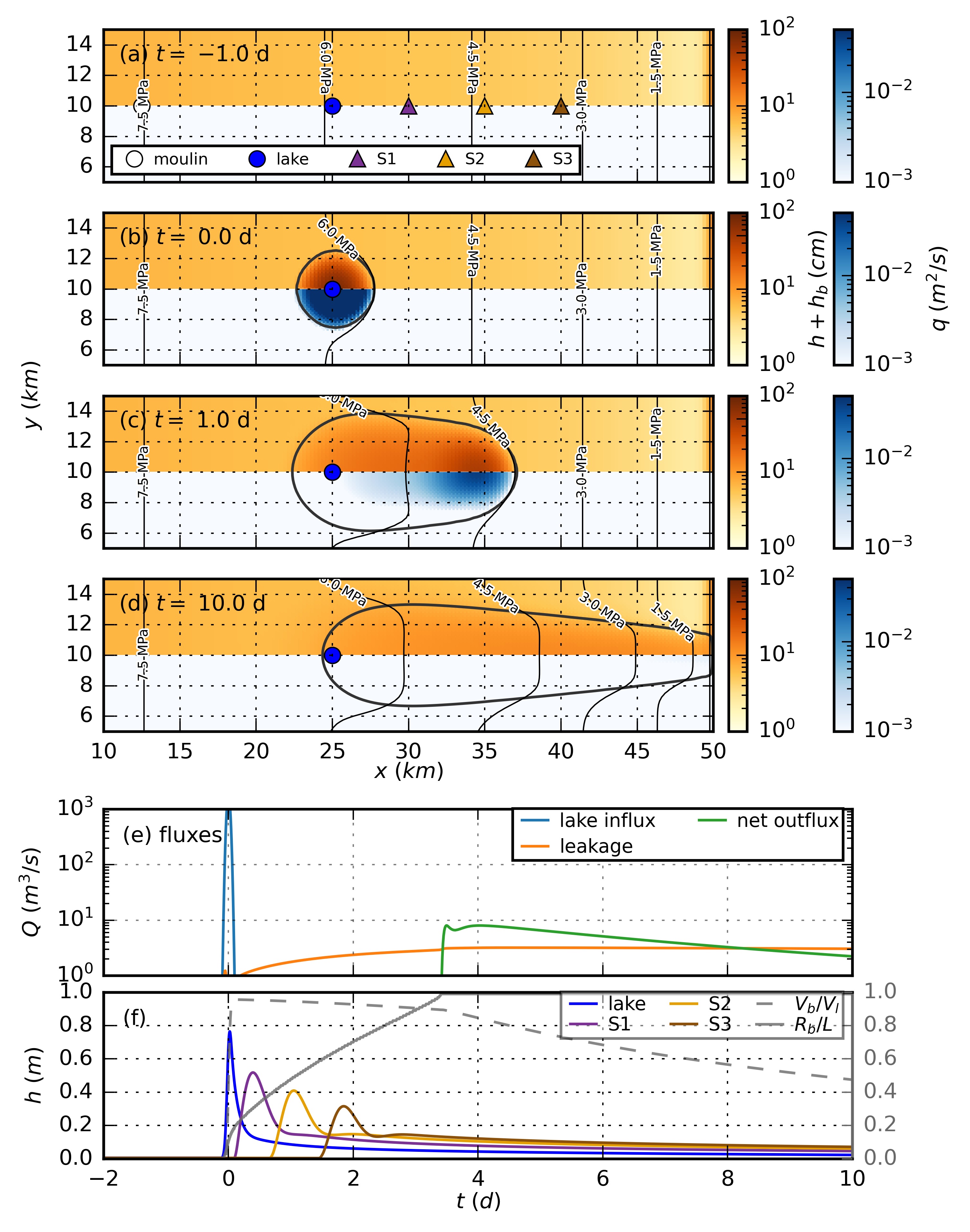}
    \caption{ \textbf{Total water thickness and flux profiles following a wintertime lake-drainage event.} \textbf{(a)} Pre-drainage state: the upper half ($y>10$ km) shows the total water thickness, $h_b+h$, and the lower half ($y<10$ km) shows the total water flux, defined as the sum of the cavity-sheet, channel, and blister fluxes. The positions of the moulin, lake, and hypothetical monitoring stations (S1-S3) are marked with coloured dots or triangles. Thin black lines are contours of the hydraulic potential $\phi$. \textbf{(b}–\textbf{d)} Total water thickness, $h_b+h$ (upper half), and total water flux magnitude (lower half), and the $0.01$-m contour (thick black line) of the blister thickness at $t=0.0$ d, $t=1.0$ d and $t=10.0$ d.  \textbf{(e)} Time series of the fluxes. The blue line is the lake-drainage input (the moulin input is zero in this case). The green line is the net outflux from the right boundary. The orange line is the leakage from the blister into the cavity–channel drainage system. \textbf{(f)} Time series of the blister thickness at different locations along the centerline. The dashed-grey line is the ratio of the blister volume to the volume of the drained lake ($V_b/V_l$). The solid-grey line is the ratio of the distance from the lake position to the blister front, to the total distance from the lake to the downstream boundary. When the blister front reaches the downstream boundary, this ratio equals 1.}
    \label{fig:wintertime_drainage}
\end{figure}

The wintertime subglacial drainage system is characterised by inefficient linked cavities. By spinning up the model without moulin input, the pre-blister drainage system only develops a thin cavity sheet due to melt sourced from constant geothermal heat fluxes, with the average thickness of this cavity sheet being approximately $0.03$ m. When the blister is generated by the lake drainage, it starts spreading roughly radially, and then propagates downstream along the ice-bed interface driven by the ice bending stress and gravity. \figref{fig:wintertime_drainage}a-d shows the magnitude of the water fluxes and the location of the blister margin, which we define as the $h_b=0.01$ m contour, at different times over this event. The water flux is defined as the sum of the fluxes in the cavity sheet, the channels and the blister. The $\phi$-contours of the cavity-channel system (\figref{fig:wintertime_drainage}a-d) indicate where the blister leaks into the surrounding drainage system and elevates the local hydraulic potential. However, since the cavity sheet is thin and the channels are very small, the volume decrease of the blister is limited until the blister reaches the downstream, ice-margin boundary at $t\approx3$ d (\figref{fig:wintertime_drainage}e). The cavity sheet beneath the propagating blister is almost instantly saturated by leakage (i.e. the local cavity sheet is filled with water from the blister). Although this results in some initial channel growth \cite{dow2015Modeling}, the growth process is relatively slow compared with our simulation time of about 10 days. Therefore, the blister itself remains the main pathway for meltwater routing as it propagates downstream along the centerline.

The blister propagation causes surface uplift along its trajectory, which is captured by $h_b$ values (the blister thickness) at four hypothetical monitoring stations positioned downstream along the model centerline (\figref{fig:wintertime_drainage}f). The normalised blister volume and the normalised distance from the lake to the blister front are also presented in \figref{fig:wintertime_drainage}f. The average front velocity is approximately $0.1~\text{m}~\text{s}^{-1}$, which is around the middle of observed subglacial flood propagation speeds summarised in \citeA{stevens2022Tidewaterglacier}: from $10^{-2}~\text{m}~\text{s}^{-1}$ to $1.0~\text{m}~\text{s}^{-1}$. A detailed exploration of the blister velocity and geometry controlled by the parameter space is presented in \secref{sec:discussion}.

\subsection{Summertime lake drainage}\label{sec:summertime}
\begin{figure}[!ht]
    \centering
    \includegraphics[width=0.8\linewidth]{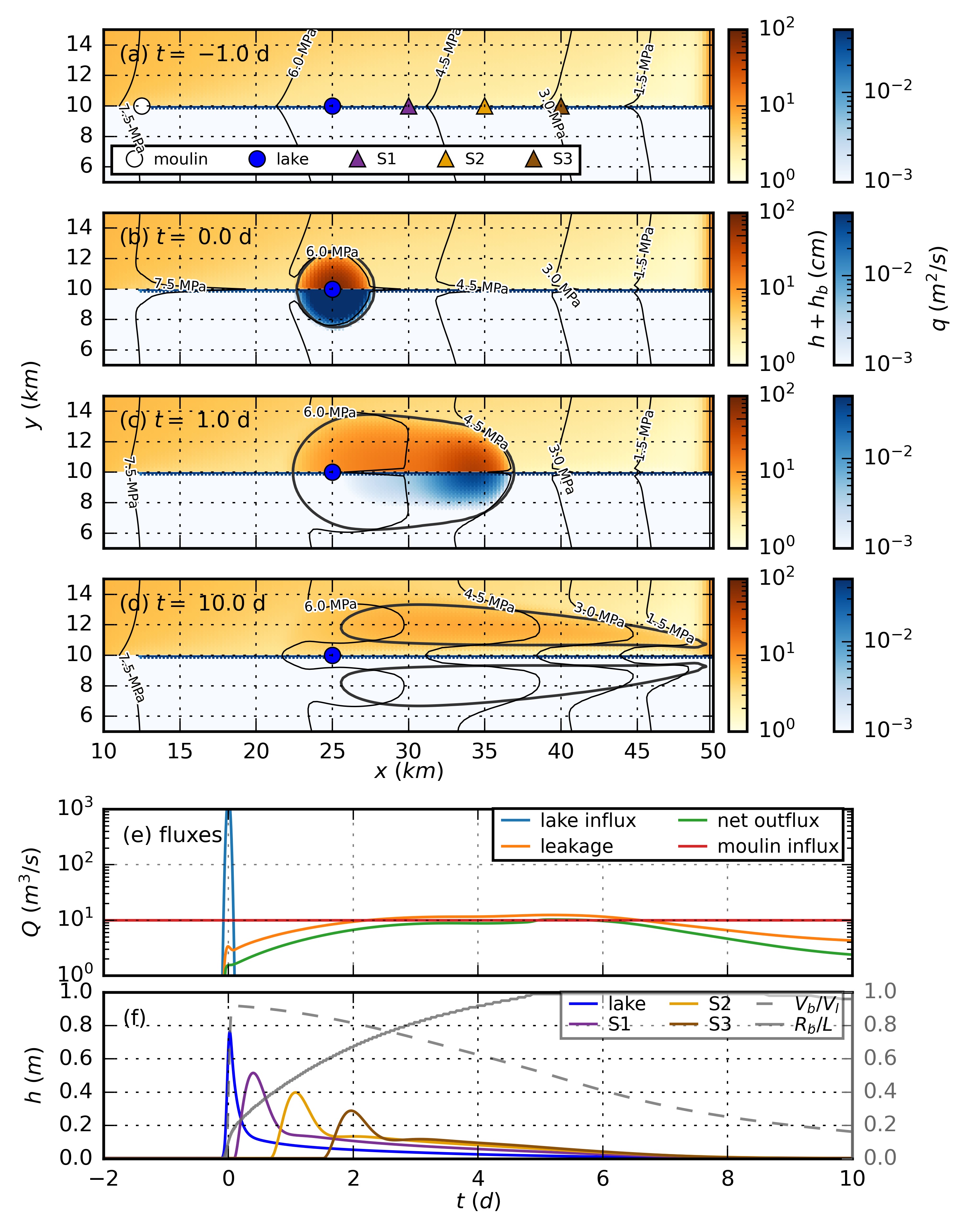}
    \caption{\textbf{Evolution of the subglacial drainage system and blister following a summertime lake drainage event.} Figure convention is the same as \figref{fig:wintertime_drainage}, except that the moulin influx ($Q_m=10~\text{m}^3~\text{s}^{-1}$), represented by the red line in (\textbf{e}), is non-zero, which leads to a subglacial channel along the centerline.}
    \label{fig:summertime_drainage}
\end{figure}

In contrast to wintertime drainage events, summertime blisters enter an evolving cavity-channel system and could rapidly leak into an efficient channel network (e.g., \citeA{lai2021Hydraulic}). For a summertime case, we use the same computational domain and parameter values (\figref{fig:wintertime_drainage}b) and spin up the model with a moulin input of $Q_m=10~\text{m}^3~\text{s}^{-1}$ (\figref{fig:summertime_drainage}a). This leads to the formation of a channel with a cross section of approximately $10~\text{m}^2$ along the ice centreline. Once the blister forms, the channel network rapidly evacuates water from the blister. This is shown by the lake-drainage flux perturbation propagating almost instantaneously along the channel (\figref{fig:summertime_drainage}e), and the blister-induced pressure perturbation remaining small near the channel (\figref{fig:summertime_drainage}b-d). Consequently, the summertime blister decays rapidly in volume (\figref{fig:summertime_drainage}f), begins vanishing from the centreline (i.e., where the channel is located), and then finally leaves two small-thickness branches on both sides of the channel (\figref{fig:summertime_drainage}d). As the blister leaks water into the channel, the channel cross-section enlarges in response to the intense meltwater input. After the blister passes, the channel cross-section decreases rapidly and eventually returns to its pre-drainage state over tens of days through viscous relaxation.

Meanwhile, as the perturbation caused by lake drainage propagates rapidly through the channel network, the blister may pressurise the channel, leading to a transient increase in water pressure at upstream locations, such as the moulin in \figref{fig:summertime_drainage}. This suggests a potential mechanism for long-range coupling within the subglacial hydrological system.


\section{Discussion}\label{sec:discussion}
The unified blister and subglacial hydrology model extends existing subglacial hydrology models by capturing transient responses to rapid lake drainage events. Previous studies, such as the one-dimensional coupling of an elastic blister and subglacial drainage system by \citeA{pimentel2010Numerical}, and the combination of an idealised hydrofracture blister model with one-dimensional subglacial hydrology models by \citeA{rice2015Time} and \citeA{dow2015Modeling}, did not account for sideways spreading or realistic downslope blister propagation. By contrast, our two-dimensional framework allows the blister to spread laterally and propagate downslope over realistic topography, enabling testable predictions of blister geometry and propagation velocity.

Our results for an idealised, land-terminating ice sheet suggest a trade-off between blister propagation and leakage into the surrounding subglacial drainage system. This trade-off controls the speed and spatial extent of surface velocity and ice uplift waves observed following supraglacial lake drainage events \citeg{maier2023Wintertime}. In this section, we examine how blister propagation velocity depends on effective viscosity $\mu_{\text{eff}}$ and lake drainage volume $V_l$ under conditions of negligible leakage. We subsequently investigate how the leakage rate is controlled by the coefficient $\kappa$, where $\kappa$ sets the efficiency with which blister water is transferred into the surrounding drainage system. This leakage analysis is conducted under conditions of slow blister propagation, with a velocity of about $5\times10^{-4}~\text{m}~\text{s}^{-1}$. Our model results are then compared with lake-drainage events reported by \citeA{maier2023Wintertime}.

\subsection{Propagation velocity}\label{sec:propagation}
Given that propagation velocity is a common observable of subglacial floods, it is important to understand the controls on propagation velocity in our model, and to examine the model's ability to propagate blisters at realistic velocities. Our two reference cases of blister propagation have front velocities of approximately $0.1~\text{m}~\text{s}^{-1}$ (\figref{fig:wintertime_drainage} \& \ref{fig:summertime_drainage}). However, velocity waves caused by rapid lake drainage events can propagate at speeds up to $1.0~\text{m}~\text{s}^{-1}$ \citeg{bjornsson2003Subglacial}. \citeA{stevens2022Tidewaterglacier} summarised a range of flood propagation speeds following, not only lake drainages, but also other types of subglacial floods such as jökulhlaups and precipitation events; they found that the propagation velocity can vary widely, from $10^{-2}~\text{m}~\text{s}^{-1}$ to $1.0~\text{m}~\text{s}^{-1}$, depending on the flood type and the local subglacial drainage system.

Here we explore how the propagation velocity in our model depends on the effective viscosity $\mu_{\text{eff}}$ and the lake drainage volume $V_l$ under the assumption of negligible leakage (i.e., $\kappa=0$). (\secref{sec:regional_study} later considers the effects of leakage on propagation in a realistic setting.) We define the front as the downstream end of the blister where the blister thickness exceeds a threshold of $10^{-2}~\text{m}$. The propagation velocity $\bar{v}$ is then defined as the average speed of the downstream ($x$-direction) front during the propagation.

\begin{figure}[!ht]
    \centering
    \includegraphics[width=1.0\linewidth]{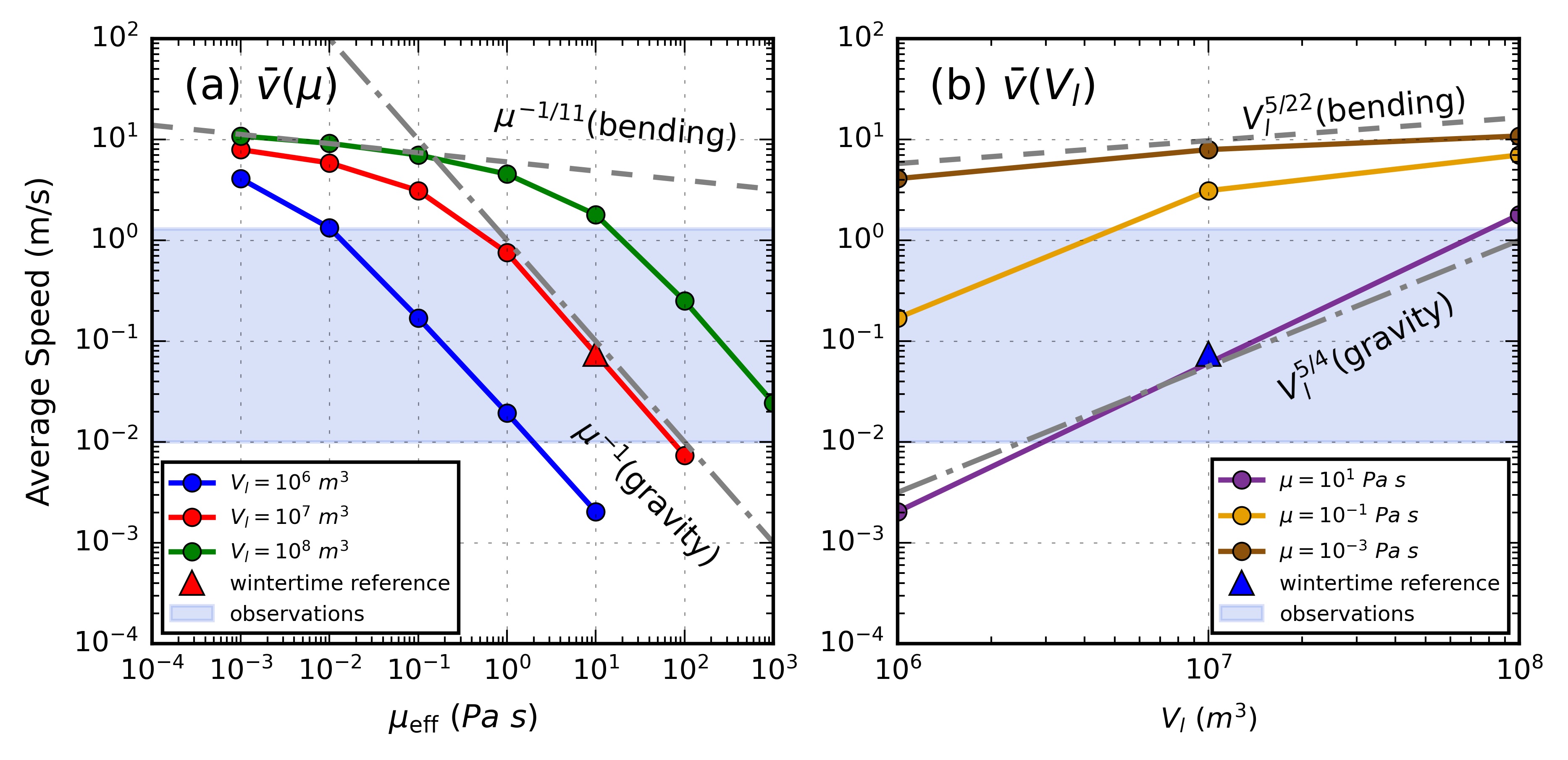}
    \caption{\textbf{Propagation velocity of the blister front as a function of effective viscosity $\mu_{\text{eff}}$ and lake drainage volume $V_l$.} (\textbf{a}) The propagation velocity $\bar{v}$ as a function of $\mu_{\text{eff}}$ for different lake drainage volumes $V_l$. The dashed line is the scaling in the bending-dominated regime ($\bar{v}\sim \mu_{\text{eff}}^{-1/11}$), while the dash-dotted line is the gravity-dominated regime ($\bar{v}\sim \mu_{\text{eff}}^{-1}$). Note that these lines just represent the trend of the dependence, and the prefactors vary among different cases and are not included. (\textbf{b}) The propagation velocity $\bar{v}$ as a function of $V_l$ for different effective viscosities $\mu_{\text{eff}}$. The dashed line and the dash-dotted line are the scalings in the bending-dominated regime ($\bar{v}\sim V_l^{5/22}$) and the gravity-dominated regime ($\bar{v}\sim V_l^{5/4}$), respectively.}
    \label{fig:blister_velocity}
\end{figure}

We vary $\mu_{\text{eff}}$ from $10^{-3}~\text{Pa}~\text{s}$ to $10^{3}~\text{Pa}~\text{s}$, covering a wide range of possible values. The lake drainage volume $V_l$ is varied from $10^{6}~\text{m}^3$ to $10^{8}~\text{m}^3$, which covers the typical range of supraglacial lake volumes in Greenland \citeg{melling2024Evaluation}. For a fixed $V_l$, the propagation velocity $\bar{v}$ decreases with increasing $\mu_{\text{eff}}$, and undergoes a transition from a bending-dominated regime to a gravity-dominated regime (\figref{fig:blister_velocity}a). The bending-dominated regime characterises the early stage of blister propagation, when the blister expands axisymmetrically, driven by the bending stresses in the overlying ice. The gravity-dominated regime generally occurs later, when the extent of the blister is larger, the bending stresses are less, and the topographically-driven hydraulic potential gradient influences the propagation. 
Previous results (\citeA{peng2020Viscous,tobin2023Evolution}; see \appref{apdx:scalings} for details) suggest that $\bar{v}\sim \mu_{\text{eff}}^{-1/11}$ during the bending-dominated regime, while $\bar{v}\sim \mu_{\text{eff}}^{-1}$ in the gravity-dominated regime.
By comparing the numerical results with the observed range of the propagation velocity, we can infer that the effective viscosity $\mu_{\text{eff}}$ likely lies between $10^{-2}~\text{Pa}~\text{s}$ and $10^{3}~\text{Pa}~\text{s}$ for typical lake drainage events with $V_l$ between $10^{6}~\text{m}^3$ and $10^{8}~\text{m}^3$, indicating that the effective viscosity is much larger than the viscosity of water ($10^{-3}~\text{Pa}~\text{s}$).

Another key factor influencing $\bar{v}$ is the blister volume. During winter, when leakage into the surrounding drainage system is negligible, the blister volume is expected to closely match the lake drainage input volume $V_l$. Variations in $V_l$ therefore directly affect the geometry of the blister, including its thickness and lateral extent, and hence modify the pressure gradients and elastic or gravitational stresses that drive propagation. As shown in \figref{fig:blister_velocity}b, lake volume exerts a clear control on $\bar{v}$. For typical lake-drainage events with $V_l$ between $10^{6}~\text{m}^3$ and $10^{8}~\text{m}^3$, the predicted propagation velocity spans several orders of magnitude, from approximately $10^{-3}~\text{m}~\text{s}^{-1}$, corresponding to an effectively stationary blister on the timescale of the drainage event, to $10^{1}~\text{m}~\text{s}^{-1}$, which is an order of magnitude faster than observed propagation speeds \cite{bjornsson2003Subglacial,stevens2022Tidewaterglacier}.

The sensitivity of $\bar{v}$ to $V_l$ also depends on the dominant force balance. For sufficiently small $\mu_{\text{eff}}$, blister propagation follows the bending-dominated regime, with $\bar{v}\sim V_l^{5/22}$ across the range of lake volumes considered. The weak exponent indicates that, in this regime, increasing lake volume produces only a modest increase in propagation velocity because the blister motion is primarily controlled by elastic bending of the overlying ice. As $\mu_{\text{eff}}$ increases, propagation slows and the system transitions toward a gravity-dominated regime, in which $\bar{v}\sim V_l^{5/4}$. In this regime, the dependence on lake volume is much stronger, indicating that larger drainage events can generate substantially faster propagation. These results suggest that both lake volume and effective viscosity are required to determine whether a lake-drainage event produces a slowly expanding, locally confined blister or a rapidly propagating subglacial water pulse.

By varying both $\mu_{\text{eff}}$ and $V_l$, our model can reproduce a wide range of blister propagation velocities that are consistent with observed subglacial flood propagation speeds. This suggests that our model effectively captures the essential physics governing blister propagation following supraglacial lake drainage events.

\subsection{Leakage}
In \secref{sec:propagation}, we explored blister propagation under the assumption of negligible leakage ($\kappa=0$). However, in reality, blister leakage into the surrounding drainage system can be significant, especially during summertime drainage events when an efficient channel network exists. Here we explore how the leakage rate depends on the leakage coefficient $\kappa$ in our ad-hoc formulation of the leakage lengthscale $l=\kappa h_b$. We consider two scenarios: a wintertime drainage event with no background moulin input ($Q_m=0$), and a summertime drainage event with a background moulin input of $Q_m=10~\text{m}^3~\text{s}^{-1}$, which leads to channel formation along the ice centreline. The effective viscosity is set to $\mu_{\text{eff}}=10^3~\text{Pa}~\text{s}$, causing the blister to move slowly and preventing it from reaching the downstream boundary during the simulation period. We vary $\kappa$ to explore its influence on the leakage behaviour.

\begin{figure}[!ht]
    \centering
    \includegraphics[width=1.0\linewidth]{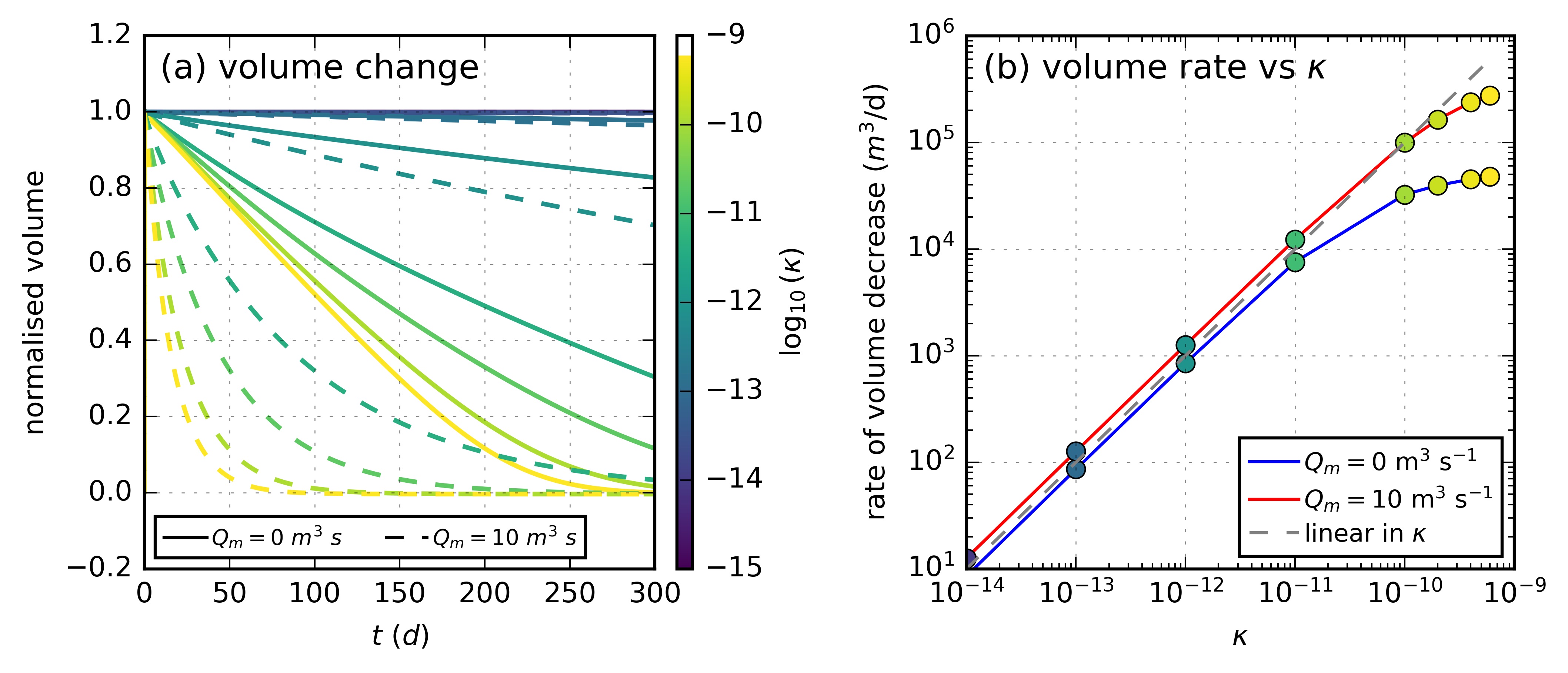}
    \caption{\textbf{Wintertime and summertime blister volume loss with different values of $\kappa$}. (\textbf{a}) Time series of the normalised blister volume $V_b/V_{l}$ due to leakage for different values of $\kappa$ in the winter case (solid lines) and the summer case (dashed lines). (\textbf{b}) The magnitude of the time-averaged volume loss rate from $t=0$ to $t=30$ d as a function of $\kappa$ for the winter case (blue) and the summer case (red). The dashed line is a linear reference. Colours correspond to different values of $\kappa$ in (\textbf{a}).}
    \label{fig:leakage}
\end{figure}

The time series of the normalised blister volume $V_b/V_l$ for different values of $\kappa$ are shown in \figref{fig:leakage}a. The volume decreases more rapidly with increasing $\kappa$, indicating a higher leakage rate. For the same $\kappa$, the summer case exhibits a higher leakage rate compared to the winter case, due to the presence of an efficient channel that quickly transports water away. However, the volume and thickness decay follow neither a simple power-law nor an exponential form \citeg{lai2021Hydraulic}, where the blister leaks into an underlying porous sheet. Here, we focus on the overall leakage rate rather than the detailed temporal evolution. To quantify this rate, we calculate the time-averaged volume-loss rate over the period from $t=0$ to $t=30$ d, during which the blister hardly moves. In both summer and winter cases, the volume-loss rate magnitude increases linearly with $\kappa$ at small values of $\kappa$, and begins to deviate from the linear trend as $\kappa$ increases (\figref{fig:leakage}b). This transition occurs when leakage reaches the capacity of the surrounding drainage system; that is, the leakage rate is limited by the ability of the cavity-channel system to transport water away, rather than the leakage process itself.

\section{A regional study in western Greenland}\label{sec:regional_study}
\begin{figure}[!ht]
    \centering
    \includegraphics[width=1.0\linewidth]{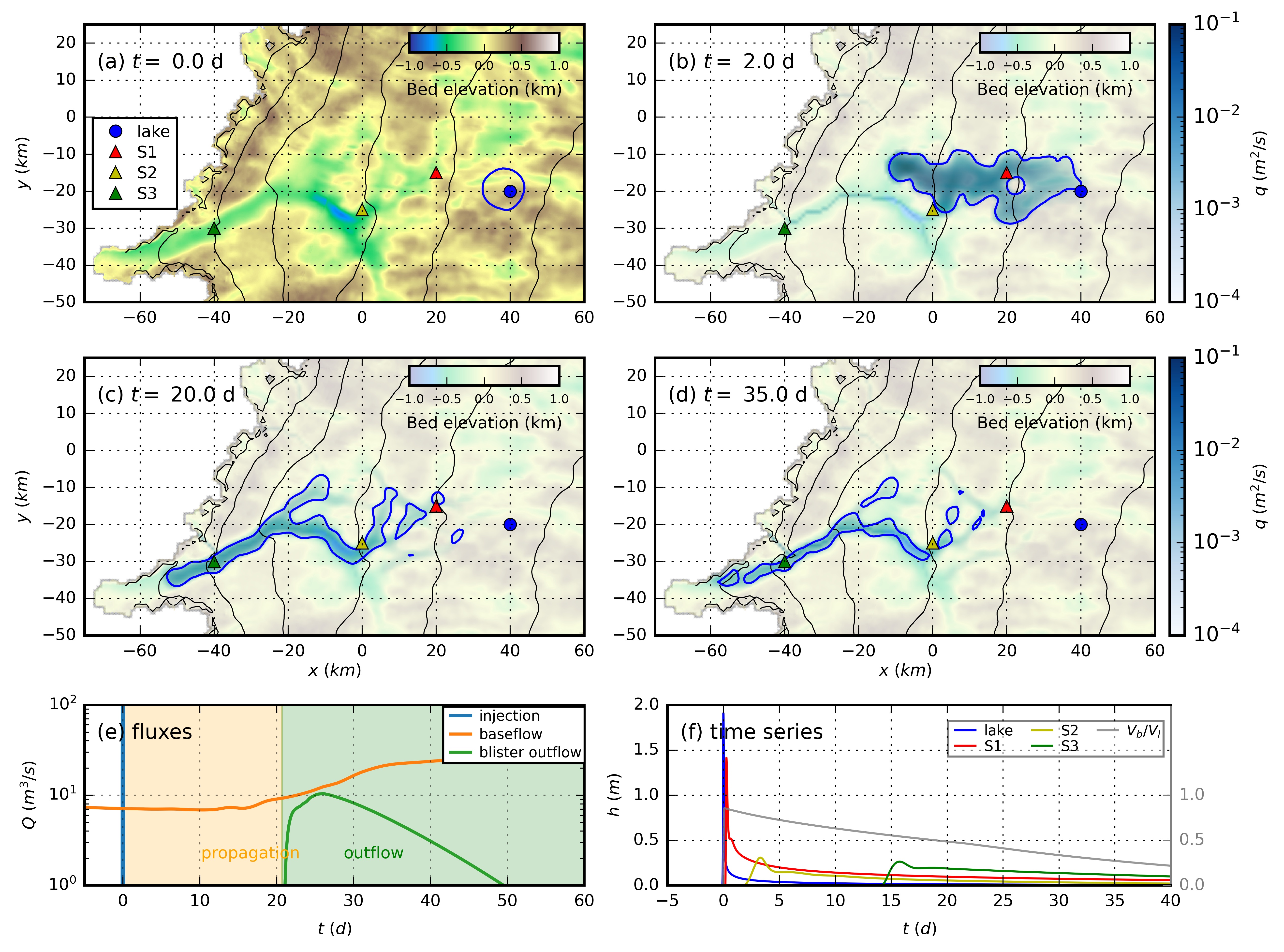}
    \caption{\textbf{Simulation of a wintertime lake drainage event in western Greenland, with $\mu_{\text{eff}}=20~\text{Pa}~\text{s}$ and $\kappa=0$.} (\textbf{a}) Bed topography and positions of the lake and hypothetical monitoring stations (S1-S3), marked with coloured dots and triangles. (\textbf{b})-(\textbf{d}) Water flux magnitude (background blue colour) and $0.1$-m contour (blue line) of the blister thickness at $t=2.0$ d, $t=20.0$ d and $t=35.0$ d. Black lines are contours of the hydraulic potential $\phi$. (\textbf{e}) Time series of the fluxes. The blue line is the lake input, with maximum $Q_{l}\sim3000~\text{m}^3~\text{s}^{-1}$. The orange line is the background water outflow of the cavity-channel system. The green line is the blister outflow. (\textbf{f}) Time series of the blister thickness at different locations along the blister path. The grey line is the ratio of the blister volume to the volume of the drained lake ($V_b/V_l$).}
    \label{fig:regional_study}
\end{figure}

To demonstrate the applicability of our model to an observed lake drainage event, we simulate an idealised scenario that mimics the wintertime lake drainage cascades documented by \citeA{maier2023Wintertime} across a land-terminating ice-sheet region in western Greenland. During the event in March 2018, a total of 18 supraglacial lakes drained sequentially, delivering approximately $1.8\times10^{8}~\text{m}^3$ of water to the ice-sheet bed. Three-component interferometric surface velocity fields and decomposition modelling indicate that the resulting subglacial flood propagated approximately $160$ km over a period of about $30$ days, causing ice-flow acceleration along the flood pathway \cite{maier2023Wintertime}. We use a $140~\text{km}\times80~\text{km}$ model domain covering the same region studied by \citeA{maier2023Wintertime}. Following \citeA{stevens2018Relationship}, the bed topography and ice surface elevation are extracted from IceBridge BedMachine Greenland Version 2 \cite{morlighem2014Deeply} and the Greenland Ice Mapping Project (GIMP) digital elevation model \cite{howat2014Greenland}, respectively. Ice thickness is calculated as the difference between ice surface elevation and bed elevation. The model domain and bed topography are shown in \figref{fig:regional_study}a.

We assume zero moulin input ($Q_m=0$) for the wintertime drainage conditions. The model is spun up for $5$ years to allow the cavity-channel system to reach a steady state driven by basal melt from constant geothermal heat fluxes and basal frictional heating. Due to the regularisation parameter $h_0=0.005~\text{m}$ in \eqnref{eq:mass_conservation_blister}, the blister layer $h_b$, initially set to zero, develops small, but non-zero, thickness during the spin-up period due to hydraulic potential gradients resulting from ice overburden and bed topography. This results in minor water redistribution within the blister layer even prior to lake drainage. However, the blister thickness remains sufficiently small to have negligible influence on the cavity-channel system and subsequent subglacial flood dynamics following lake drainage. A more physically consistent regularisation scheme, such as incorporating fracture mechanics \cite{lister2019Viscous}, could resolve this issue, but this is left to future work. Additionally, we idealise the lake-drainage event as a single-point input at $(x,y)=(40~\text{km},-20~\text{km})$, delivering a total water volume of $1.8\times10^{8}~\text{m}^3$ over a timescale of $\tau_l=0.6$ hr. 

We first demonstrate the blister behaviour by considering a case with zero leakage ($\kappa=0$). The observed propagation velocity of approximately \(0.1~\mathrm{m~s^{-1}}\) is used to calibrate \(\mu_{\mathrm{eff}}\), giving a representative value of \(\mu_{\mathrm{eff}}=20~\mathrm{Pa~s}\), rather than prescribing \(\mu_{\mathrm{eff}}\) independently. On realistic topography, the blister does not propagate along a straight line as in the idealised cases in \secref{sec:results}. Instead, it propagates toward the ice margin following a tortuous path determined by a combination of ice overburden, bed topography, and ice bending stress. In \figref{fig:regional_study}a-d, we present both the blister water flux magnitude and the blister $0.1$-m thickness contour at various times. The blister reaches the ice margin in $25$ days (\figref{fig:regional_study}d-e), corresponding to an average propagation velocity of approximately $0.1~\text{m}~\text{s}^{-1}$. 

The $h_b$ time series at the lake location and at the three downstream, hypothetical monitoring stations show diverse patterns of surface uplift (\figref{fig:regional_study}f). When the drainage occurs, the lake location and station S1 both experience uplift of over $1$ m, which subsequently decays. Other downstream locations experience smaller uplifts with relatively slower decay rates. Notably, due to the presence of the blister's tail region, uplift at these downstream locations persists long after the blister front has passed, which could help explain the prolonged uplift signals observed after lake drainage events by \citeA{doyle2013Ice} and \citeA{mejia2021Isolated}.

\begin{figure}[!ht]
    \centering
    \includegraphics[width=1.0\linewidth]{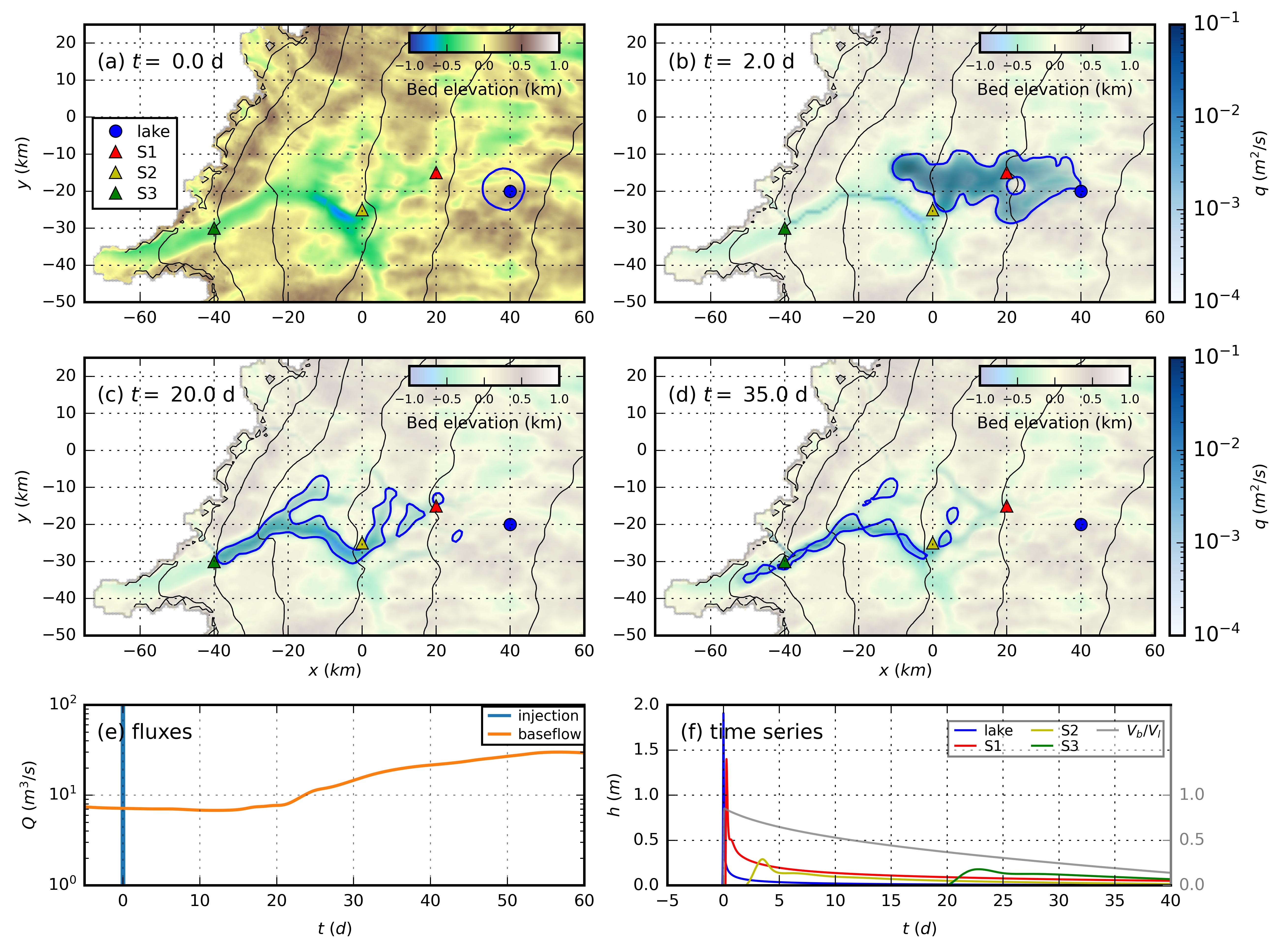}
    \caption{\textbf{Simulation of a wintertime lake drainage event in western Greenland, with $\mu_{\text{eff}}=10~\text{Pa}~\text{s}$ and $\kappa=10^{-11}$.} Figure convention is the same as \figref{fig:regional_study}.}
    \label{fig:regional_study_leakage}
\end{figure}

Since leakage, which is parameterised by $\kappa$, governs the rate of blister volume loss, leakage may be as critical as blister propagation in controlling subglacial meltwater routing and surface uplift patterns. To explore the effects of leakage, we conduct a series of simulations using $\mu_{\text{eff}}=10~\text{Pa}~\text{s}$ and various values of $\kappa\ge 0$. A representative case ($\kappa=10^{-11}$) is shown in \figref{fig:regional_study_leakage}. Due to the leakage, and consequent blister deceleration during propagation, the blister does not reach the ice margin when $\kappa=10^{-11}$; instead, blister water moves into the cavity sheet and channels through leakage, contributing to an increase in baseflow.

\begin{figure}[!ht]
    \centering
    \includegraphics[width=0.85\linewidth]{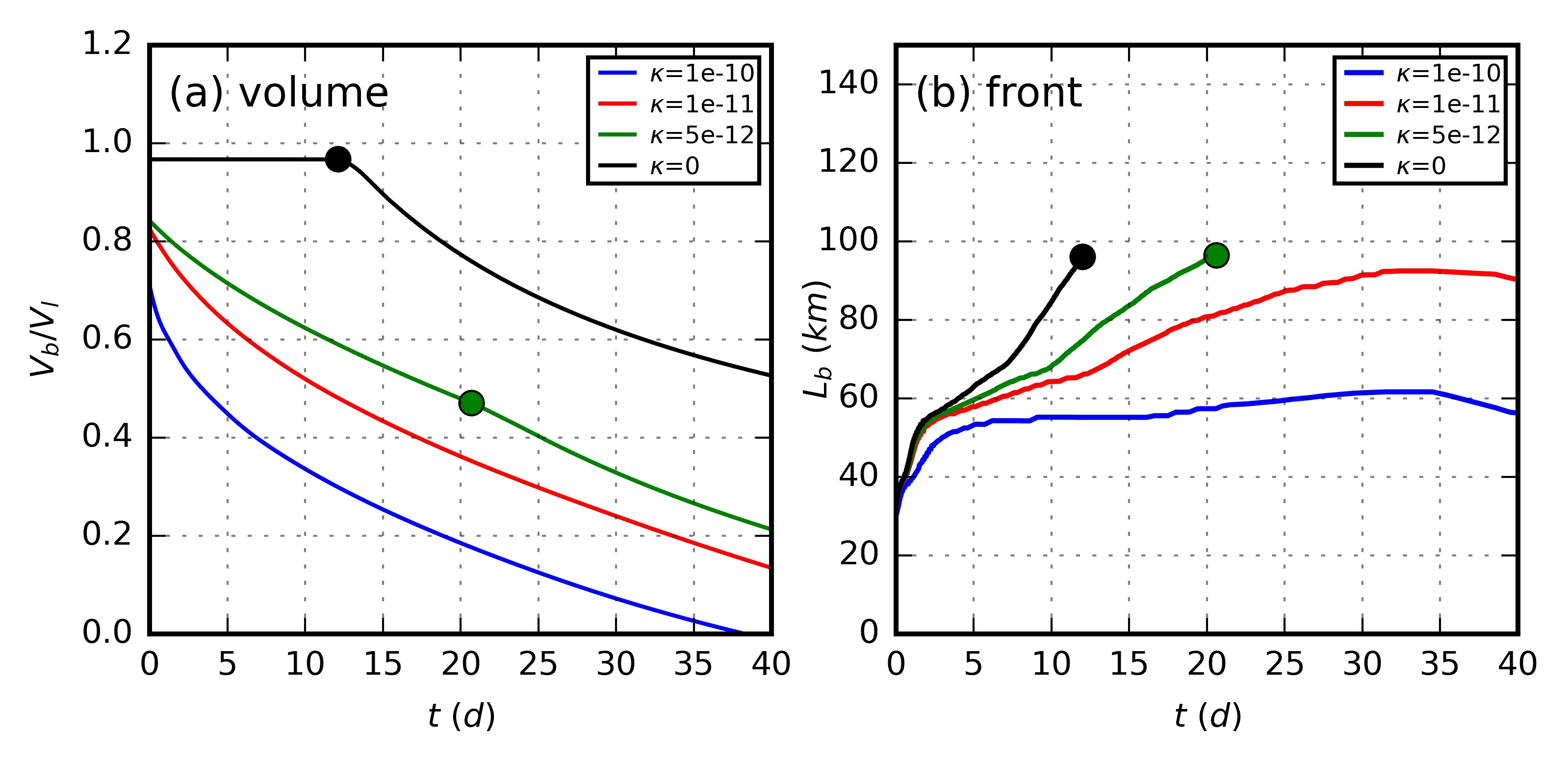}
    \caption{\textbf{Time series of the blister volume and front position in the regional study with different values of $\kappa$.} (\textbf{a}) Time series of the normalised blister volume $V_b/V_{l}$, and (\textbf{b}) distance from the lake to the blister front $L_b$, for different values of $\kappa$ with $\mu_{\text{eff}}=10~\text{Pa}~\text{s}$. Dots indicate the time when the blister front reaches the ice margin, with colours representing different values of $\kappa$. No dots are shown for cases where the blister front does not reach the ice margin. The red line ($\kappa=10^{-11}$) represents the case shown in \figref{fig:regional_study_leakage}. In the supplementary material, we provide videos illustrating the blister propagation and leakage for all four different values of $\kappa$.}
    \label{fig:regional_study_time_series}
\end{figure}

Results of other cases, from $\kappa=0$ to $\kappa=10^{-10}$, are summarised in \figref{fig:regional_study_time_series}. For simplicity, we define the blister front position as the westernmost point of the blister where its thickness exceeds $0.1$ m. As indicated by \eqnref{eq:leakage}, the blister volume decreases more rapidly with increasing $\kappa$ (\figref{fig:regional_study_time_series}a). In \figref{fig:regional_study_time_series}b, we plot the distance from the lake to the blister front, $L_b$, as a function of time. As leakage increases (i.e., with larger $\kappa$), the blister front propagates more slowly due to rapid volume loss, and propagation may eventually halt before reaching the ice margin if $\kappa$ becomes sufficiently large. The drained water is then primarily routed through the channel network. By combining observed propagation speeds, surface uplift patterns, and water outflow locations and timings at the ice margin, this model can potentially constrain the values of $\mu_{\text{eff}}$ and $\kappa$ for specific lake drainage events, thereby determining the primary pathways of subglacial flood routing. 

\section{Limitations}\label{sec:limitations}
As a first step towards modelling blister–hydrology interactions, the framework illustrates how blisters may interact with cavity sheets and channels. However, the use of the laminar flow law with an effective viscosity and the parameterisation of leakage are both somewhat ad hoc. Future theoretical and numerical studies could adopt a turbulent blister model, and focus in more detail on how blisters promote channel growth and interact with surrounding cavities.

Another key limitation of the present model is that it does not resolve the hydrofracture process that initiates rapid lake drainage and delivers surface water to the bed. Instead, lake drainage is prescribed as an observationally supported input flux \cite{das2008Fracture,doyle2013Ice,stevens2015Greenland,chudley2019lake}, and the model focuses on the subsequent redistribution of water through the coupled blister and subglacial drainage system. Recent work has shown that ice viscosity can strongly influence hydraulic-fracture propagation during supraglacial lake-drainage events \cite{hageman2024Ice}, highlighting the importance of considering viscous ice deformation during the initial lake-drainage process. 

Finally, the blister in our framework is treated as water flow beneath an elastically bending ice sheet. This elastic approximation is appropriate for representing the short-term, flexural response of the overlying ice after water has reached the bed, but it neglects viscous deformation of the ice during blister relaxation. A viscous \citeg{evatt2007Cauldron} or viscoelastic \citeg{walker2013Iceshelf} beam model would provide a more realistic description of long-term relaxation driven by ice creep. Future developments should couple the present subglacial hydrology framework to a vertical hydrofracture model and incorporate viscoelastic ice bending. Such extensions would enable more realistic simulations of both the initiation of lake-drainage events and the subsequent evolution of surface uplift.

\section{Conclusion}\label{sec:conclusion}
In this study, we have developed a unified model that couples blister dynamics with the subglacial drainage system, allowing us to explore the transient effects of rapid supraglacial lake drainage events. Our model captures the seasonal variability of blister dynamics, demonstrating that summertime drainage leads to efficient blister leakage into existing channel networks, while wintertime drainage results in persistent blisters that act as primary pathways for meltwater transport. The dynamics of blister propagation and leakage in our model are governed by effective viscosity and a characteristic leakage length scale. An exploration of the parameter space reveals that blister propagation velocity can vary significantly, depending on the effective viscosity and the volume of the lake drainage input.

This integrated model enables us to investigate blister dynamics and how blister formation and propagation interact with, and modify, subglacial hydrology under realistic ice-sheet conditions. Our regional simulation of a wintertime lake drainage event in western Greenland produces results consistent with observed propagation speeds and subglacial flood extents. The model can facilitate the interpretation of observed surface uplift and ice-velocity anomalies following supraglacial lake drainage events. Future work will focus on developing consistent formulations of effective viscosity and leakage, and coupling the current model to an ice-sheet sliding model to better capture feedbacks between ice dynamics and subglacial hydrology.

\appendix

\section{Numerical implementation and convergence test}\label{apdx:convergence_test}
The model is implemented using the finite volume method with a staggered grid as described in \citeA{hewitt2013Seasonal}. The channels are defined on the edge of the grid cells, while the cavity sheet and blister are defined at the centre of the grid cells. Details of the numerical implementation of the cavity-channel system can be found in \citeA{hewitt2013Seasonal}.

The blister dynamics \eqnref{eq:hydraulic_potential_blister} and \eqnref{eq:mass_conservation_blister} are solved using an implicit scheme for time stepping. The spatial derivatives are discretised using second-order central difference. The blister equations, together with the cavity-channel equations, form a coupled nonlinear system that is solved using the Newton-Raphson method at each time step. Here we present convergence tests for both bending-dominated and gravity-dominated regimes to verify the mesh-independence of the numerical solution.
\begin{figure}[!ht]
    \centering
    \includegraphics[width=1.0\linewidth]{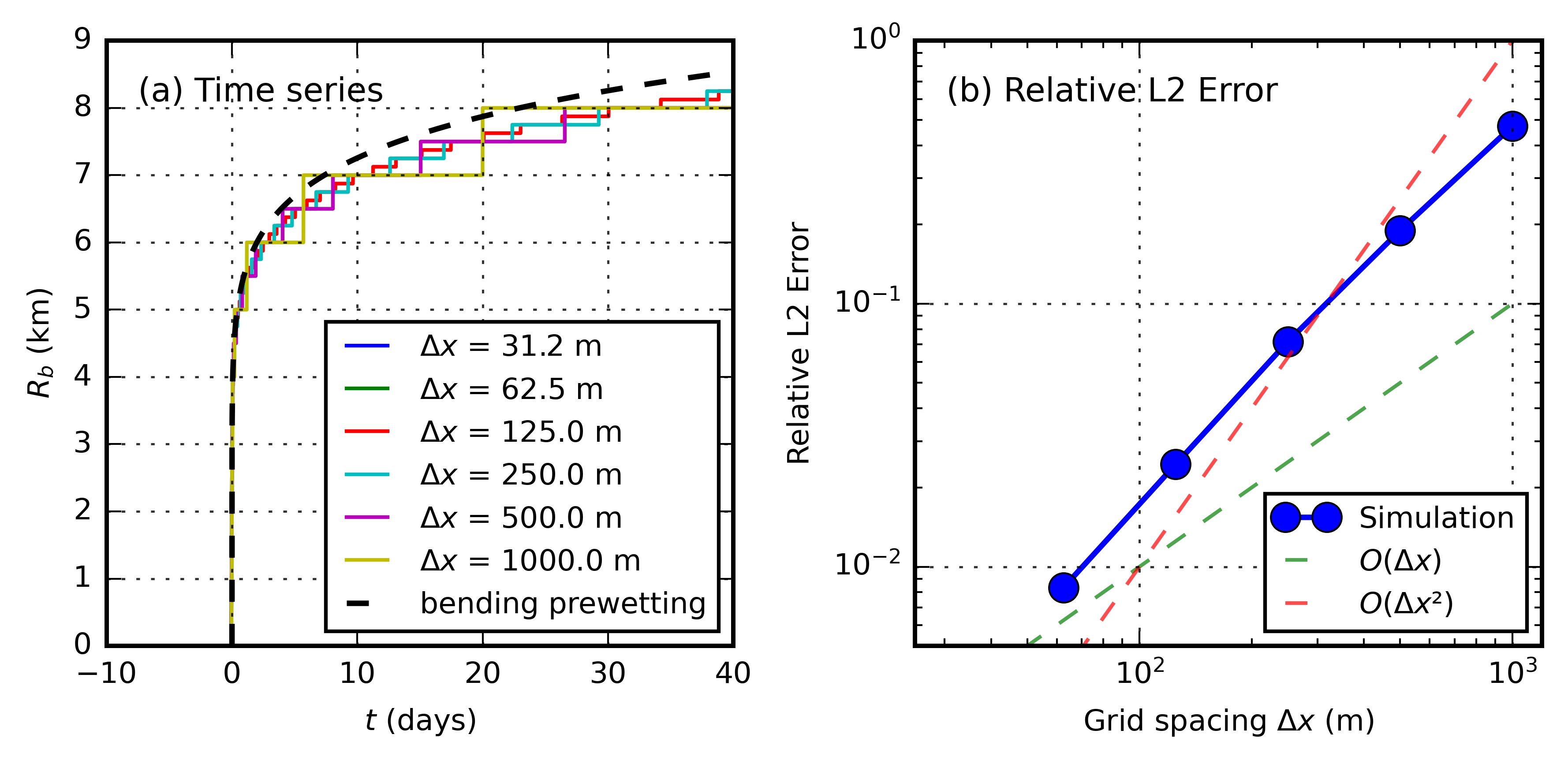}
    \caption{\textbf{Convergence test under a bending-dominated regime.} (\textbf{a}) Time series of the blister front position for different grid sizes $\Delta x$ in the bending-dominated regime. The reference solution is chosen to be the numerical solution with $\Delta x=31.25~\text{m}$. The dashed line is the scaling \eqnref{eq:rb_bending_1d}. (\textbf{b}) The relative error of the blister front position compared to the reference solution as a function of $\Delta x$.}
    \label{fig:convergence_bending_test}
\end{figure}

We have verified that the numerical solution for blister propagation converges as the grid size $\Delta x\rightarrow 0$ by simulating both bending-dominated and gravity-dominated regimes in a one-dimensional model at various grid resolutions. For the bending-dominated regime, we consider an ice sheet of uniform thickness $H=1000~\text{m}$ resting on a flat bed ($b=0$). A blister with a volume of $10^{2}~\text{m}^2$ and an effective viscosity of $\mu_{\text{eff}}=10^{-3}~\text{Pa}~\text{s}$ propagates symmetrically in both directions. We denote the blister front position as $R_b$, which is defined as the distance from the blister centre to the point where the blister thickness drops below $2h_0$.

For the gravity-dominated regime, the ice sheet has the same uniform thickness ($H=1000~\text{m}$) but rests on a linear bedrock slope of angle $\theta=10^{-2}$. The blister has the same volume but a larger effective viscosity of $\mu_{\text{eff}}=10^{1}~\text{Pa}~\text{s}$. In this case, the blister still propagates symmetrically at early times under bending stresses, but eventually shifts to a unidirectional downslope propagation, and we again use $R_b$ to denote the distance travelled by the blister front.

The grid size $\Delta x$ ranges from $1000~\text{m}$ down to $31.25~\text{m}$ in both regimes. The numerical solution obtained at the finest grid resolution, denoted as $R_{b,\text{ref}}$, serves as the reference solution. For a given grid size $\Delta x$, the relative error $E$ of the time series $R_b(t)$ is defined as 
\begin{linenomath*}
\begin{equation}\label{eq:error}
    E\left(R_b\right) = \frac{\Vert R_b\left(t\right)-R_{b,\text{ref}}(t)\Vert}{\Vert R_{b,\text{ref}}(t) \Vert},
\end{equation}
\end{linenomath*}
where $t$ ranges from $0$ to $40$ days, and $\Vert\cdot\Vert$ denotes the $L^2$ norm over time.

\begin{figure}[!ht]
    \centering
    \includegraphics[width=1.0\linewidth]{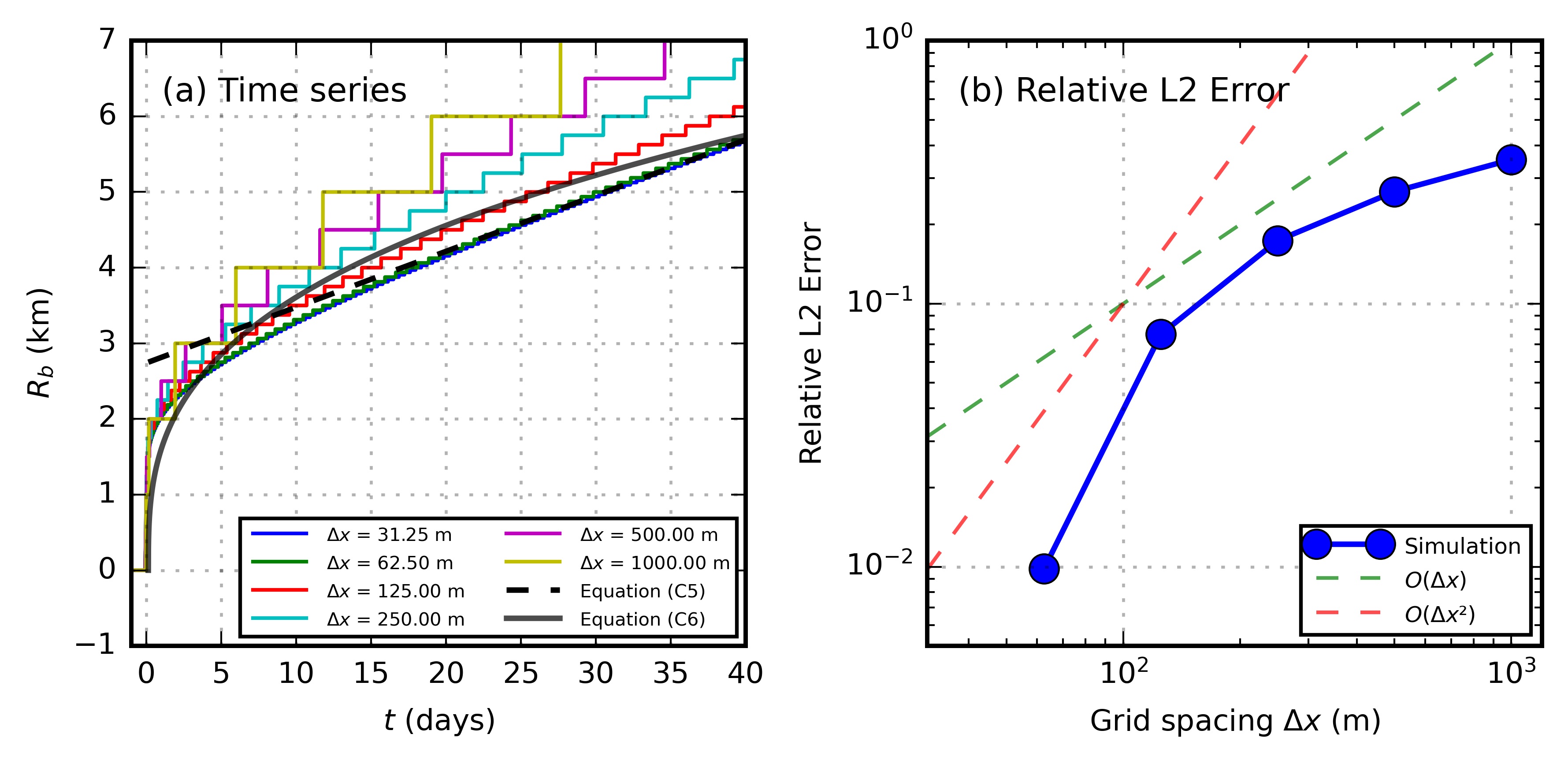}
    \caption{\textbf{Convergence test under a gravity-dominated regime.} (\textbf{a}) Time series of the blister front position for different grid sizes $\Delta x$ in the gravity-dominated regime with a linear bedrock profile at an angle of $\theta=10^{-2}$. The reference solution is chosen to be the numerical solution with $\Delta x=31.25~\text{m}$. The solid black line and the dashed black line are scalings obtained from the scaling for a gravity-dominated blister (\eqnref{eq:vel_gravity_fracture}) and the scaling for a viscous gravity current (\eqnref{eq:vel_gravity}), respectively. (\textbf{b}) The relative error of the blister front position compared to the reference solution as a function of $\Delta x$.}
    \label{fig:convergence_gravity_test}
\end{figure}

The convergence test for the bending-dominated regime is shown in \figref{fig:convergence_bending_test}. The relative error of the blister front position compared to the reference solution decreases as the grid size $\Delta x$ is reduced. The corresponding convergence test for the gravity-dominated regime is presented in \figref{fig:convergence_gravity_test}, where the relative error similarly decreases with decreasing $\Delta x$. Based on these results, we select a grid size of $\Delta x=125~\text{m}$ for the reference cases presented in \secref{sec:results}, and $\Delta x=900~\text{m}$ for the regional study in \secref{sec:regional_study}, as this grid-size choice provides a balance between accuracy and computational efficiency.

\section{Scalings for the blister propagation}\label{apdx:scalings}
Here we summarise the expected scalings for blister propagation from the literature. We start with two-dimensional, symmetric blisters with a constant volume $V_b$. The position of the blister front (equivalent to radius for two-dimensional blisters) is denoted as $R_b(t)$.

For the bending-dominated regime \cite{peng2020Viscous}, the propagation velocity $\dot{R_b}$ can be derived as
\begin{linenomath*}
\begin{equation}\label{eq:vel_bending}
    \dot{R_b} = 76.1 \frac{B h_0^{1/2} V_b^{5/2}}{12\mu_{\text{eff}}R_b^{10}},
\end{equation}
\end{linenomath*}
where $B$ is the bending stiffness, and $h_0$ is the pre-wetted layer thickness. Assuming that $R_b=0$ at $t=0$, integrating \eqnref{eq:vel_bending} over $t$ gives the blister radius as
\begin{linenomath*}
\begin{equation}\label{eq:rb_bending}
    R_b = 1.47 \left(\frac{B h_0^{1/2} V_b^{5/2}}{\mu_{\text{eff}}}\right)^{1/11} t^{1/11},
\end{equation}
\end{linenomath*}
which indicates that the dependencies of $\dot{R_b}$ on the effective viscosity $\mu_{\text{eff}}$ and the lake drainage volume $V_l$ are relatively weak, with $\dot{R_b}\sim \mu_{\text{eff}}^{-1/11}$ and $\dot{R_b}\sim V_b^{5/22}$ (\figref{fig:blister_velocity}). 

In the one-dimensional geometry used in the convergence test (\appref{apdx:convergence_test}), we use $V_b$ to denote the blister volume per unit width in the cross-slope direction. The scaling for the velocity is
\begin{linenomath*}
\begin{equation}\label{eq:vel_bending_1d}
    \dot{R_b} = 411 \frac{B h_0^{1/2} V_b^{5/2}}{12\mu_{\text{eff}}R_b^{15/2}},
\end{equation}
\end{linenomath*}
and thus for the radius,
\begin{linenomath*}
\begin{equation}\label{eq:rb_bending_1d}
    R_b = 1.95 \left(\frac{B h_0^{1/2} V_b^{5/2}}{\mu_{\text{eff}}}\right)^{2/17} t^{2/17}.
\end{equation}
\end{linenomath*}

For the gravity-dominated regime \cite{tobin2023Evolution}, propagation of the blister can be approximated as one-dimensional (along the downslope direction). In this regime, the front velocity $\dot{R_b}$ is given by
\begin{linenomath*}
\begin{equation}\label{eq:vel_gravity}
    \dot{R_b} = \frac{\left(\rho_w g V_b \sin{\alpha}\right)^{5/4} h_0^{1/2}}{12\mu_{\text{eff}} B^{1/4}},
\end{equation}
\end{linenomath*}
where $\alpha$ is the angle of the bed slope. This scaling indicates that the propagation velocity is constant in time given a constant bed slope. Note that this scaling assumes the blister has propagated sufficiently far downslope, as gravity effects need to dominate over bending effects. The dependence of $\dot{R_b}$ on the effective viscosity $\mu_{\text{eff}}$ and the lake drainage volume $V_l$ is relatively strong, with $\dot{R_b}\sim \mu_{\text{eff}}^{-1}$ and $\dot{R_b}\sim V_b^{5/4}$ (\figref{fig:blister_velocity}).

Note that our blister model with a pre-wetted layer depends on the regularisation parameter $h_0$.  Alternatively, assuming that the blister is a viscous gravity current propagating on a dry bed \citef{huppert1982Flow}, the scaling of the front speed $\dot{R_b}$ follows a power-law decay controlled by the balance between the downslope component of gravity and viscous forces:
\begin{linenomath*}
\begin{equation}\label{eq:vel_gravity_fracture}
    \dot{R_b} = \frac{1}{2^{2/3}}\left(\frac{\rho_w g V_b^2 \sin{\alpha}}{12\mu_{\text{eff}}}\right)^{1/3}t^{-2/3},
\end{equation}
\end{linenomath*}
thus
\begin{linenomath*}
\begin{equation}\label{eq:rb_gravity_fracture}
    R_b = \frac{3}{2^{2/3}}\left(\frac{\rho_w g V_b^2 \sin{\alpha}}{12\mu_{\text{eff}}}\right)^{1/3}t^{1/3}. 
\end{equation}
\end{linenomath*}

Different from the gravity-dominated blister with a pre-wetted layer, the propagation velocity of a gravity current decays with time (\eqnref{eq:vel_gravity_fracture}). In \figref{fig:convergence_gravity_test}a, we compare the front position $R_b$ of the gravity-dominated blister with a pre-wetted layer to the scaling in \eqnref{eq:vel_gravity} (integrated over time and shifted to match the numerical results at late times) and the scaling in \eqnref{eq:rb_gravity_fracture}. With the choice of $h_0=10^{-3}$ m, the numerically-simulated propagation velocity remains close to the scaling in \eqnref{eq:vel_gravity_fracture} within the simulation time of $40$ days, which justifies our choice of $h_0$ in this study.

%
%
\section*{Open Research Section}
The code used to run the simulations in this study is available at \url{https://zenodo.org/records/20523668}, which is a modified version of the subglacial hydrology model by \citeA{hewitt2013Seasonal}. The data used in the regional study in \secref{sec:regional_study} can be downloaded from the following sources: the IceBridge BedMachine Greenland Version 2 \cite{morlighem2014Deeply} at \url{https://nsidc.org/data/idbmg4/versions/2}; the Greenland Ice Mapping Project (GIMP) digital elevation model \cite{howat2014Greenland} at \url{https://nsidc.org/data/nsidc-0645/versions/1}, used by \citeA{stevens2018Relationship} at \url{https://zenodo.org/records/1299945}.

\section*{Conflict of Interest declaration}
The authors declare there are no conflicts of interest for this manuscript.

\section*{Copyright}
For the purpose of open access, the authors have applied a CC BY public copyright licence to any author accepted manuscript arising from this submission.

\acknowledgments
This work was supported by the Natural Environment Research Council (NERC) grant NE/Y002369/1, as a part of the joint NSF-NERC project ``Understanding surface-to-bed meltwater pathways across the Greenland Ice Sheet using machine-learning and physics-based models." The authors thank project members F. Clerc, C.-Y. Lai, E. Lee, J. Rines, M. Shahin, and L. Stearns for helpful discussions.

\bibliography{zhang2025subglacial}

\end{document}